\documentclass[a4paper,10pt,twoside]{cpc-hepnp}

\usepackage{subfigure}
\usepackage{multicol}
\usepackage{units}
\usepackage{booktabs}
\usepackage{amsmath}
\usepackage{amsfonts,amssymb,bm,mathrsfs,bbm,amscd}
\usepackage{lastpage}
\usepackage{multirow}
\usepackage{stfloats}
\usepackage{graphicx}
\DeclareGraphicsExtensions{.pdf,.jpeg,.png}
\usepackage{epstopdf}    
\usepackage[english]{babel}
\usepackage{overpic}
\usepackage[title]{appendix}
\usepackage{lineno}
\usepackage{color}
\usepackage{diagbox}
\usepackage{epsfig}
\usepackage{bigstrut}
\usepackage{overpic}
\usepackage{dcolumn}
\usepackage{xspace}
\usepackage{verbatim}
\usepackage{footmisc}
\usepackage{xcolor}
\usepackage{soul}

\usepackage{caption}
\usepackage{float}
\usepackage{ulem}

\usepackage[colorlinks,linkcolor=blue,anchorcolor=blue,citecolor=blue]{hyperref}

 \hypersetup{
    colorlinks=true,
    linkcolor=blue,
    citecolor=green,
    urlcolor=blue,
    pdfborder={0 0 0},
}

\newcolumntype{d}[1]{D{.}{.}{#1}}

\begin{document}



\fancyhead[c]{\small Chinese Physics C~~~Vol. xx, No. x (2024) xxxxxx}
\fancyfoot[C]{\small 010201-\thepage}
\footnotetext[0]{Received xxxx December xxxx}

\title{Determination of the number of \texorpdfstring{\(\psi(3686)\)}{Lg} events taken at BESIII \thanks{
The BESIII Collaboration thanks the staff of BEPCII and the IHEP computing center for their strong support. This work is supported in part by National Key R\&D Program of China under Contracts Nos. 2020YFA0406300, 2020YFA0406400; National Natural Science Foundation of China (NSFC) under Contracts Nos. 12150004, 11635010, 11735014, 11835012, 11935015, 11935016, 11935018, 11961141012, 12025502, 12035009, 12035013, 12061131003, 12192260, 12192261, 12192262, 12192263, 12192264, 12192265, 12221005, 12225509, 12235017;
the Program of Science and Technology Development Plan of Jilin Province of China under Contract No. 20210508047RQ and 20230101021JC;
the Chinese Academy of Sciences (CAS) Large-Scale Scientific Facility Program; the CAS Center for Excellence in Particle Physics (CCEPP); Joint Large-Scale Scientific Facility Funds of the NSFC and CAS under Contract No. U1832207; CAS Key Research Program of Frontier Sciences under Contracts Nos. QYZDJ-SSW-SLH003, QYZDJ-SSW-SLH040; 100 Talents Program of CAS; The Institute of Nuclear and Particle Physics (INPAC) and Shanghai Key Laboratory for Particle Physics and Cosmology; European Union's Horizon 2020 research and innovation programme under Marie Sklodowska-Curie grant agreement under Contract No. 894790; German Research Foundation DFG under Contracts Nos. 455635585, Collaborative Research Center CRC 1044, FOR5327, GRK 2149; Istituto Nazionale di Fisica Nucleare, Italy; Ministry of Development of Turkey under Contract No. DPT2006K-120470; National Research Foundation of Korea under Contract No. NRF-2022R1A2C1092335; National Science and Technology fund of Mongolia; National Science Research and Innovation Fund (NSRF) via the Program Management Unit for Human Resources \& Institutional Development, Research and Innovation of Thailand under Contract No. B16F640076; Polish National Science Centre under Contract No. 2019/35/O/ST2/02907; The Swedish Research Council; U. S. Department of Energy under Contract No. DE-FG02-05ER41374.}}

\maketitle
\begin{center}
\begin{small}
	\begin{center}
		M.~Ablikim$^{1}$, M.~N.~Achasov$^{4,c}$, P.~Adlarson$^{75}$, O.~Afedulidis$^{3}$, X.~C.~Ai$^{80}$, R.~Aliberti$^{35}$, A.~Amoroso$^{74A,74C}$, Q.~An$^{71,58,a}$, Y.~Bai$^{57}$, O.~Bakina$^{36}$
% ...
		\\
		\vspace{0.2cm}
		(BESIII Collaboration)\\
		\vspace{0.2cm} {\it
			$^{1}$ Institute of High Energy Physics, Beijing 100049, People's Republic of China\\
% ...
$^{3}$ Bochum Ruhr-University, D-44780 Bochum, Germany\\
			$^{4}$ Budker Institute of Nuclear Physics SB RAS (BINP), Novosibirsk 630090, Russia\\
% ...
$^{35}$ Johannes Gutenberg University of Mainz, Johann-Joachim-Becher-Weg 45, D-55099 Mainz, Germany\\
			$^{36}$ Joint Institute for Nuclear Research, 141980 Dubna, Moscow region, Russia\\
% ...
$^{57}$ Southeast University, Nanjing 211100, People's Republic of China\\
			$^{58}$ State Key Laboratory of Particle Detection and Electronics, Beijing 100049, Hefei 230026, People's Republic of China\\
% ...
$^{71}$ University of Science and Technology of China, Hefei 230026, People's Republic of China\\
% ...
$^{74}$ University of Turin and INFN, (A)University of Turin, I-10125, Turin, Italy; (B)University of Eastern Piedmont, I-15121, Alessandria, Italy; (C)INFN, I-10125, Turin, Italy\\
			$^{75}$ Uppsala University, Box 516, SE-75120 Uppsala, Sweden\\
% ...
$^{80}$ Zhengzhou University, Zhengzhou 450001, People's Republic of China\\
			\vspace{0.2cm}
			$^{a}$ Deceased\\
% ...
$^{c}$ Also at the Novosibirsk State University, Novosibirsk, 630090, Russia\\
% ...
	}\end{center}	
	\vspace{0.4cm}
\end{small}

\end{center}

\begin{abstract}
  The number of $\psi(3686)$ events collected by the BESIII detector during the 2021 run period is determined to be $(2259.3\pm 11.1)\times 10^6$ by counting inclusive $\psi(3686)$ hadronic events.  The uncertainty is systematic and the statistical uncertainty is negligible.  Meanwhile, the numbers of $\psi(3686)$ events collected during the 2009 and 2012 run periods are updated to be $(107.7\pm0.6)\times 10^6$ and $(345.4\pm 2.6)\times 10^6$, respectively. Both numbers are consistent with the previous measurements within one standard deviation. The total number of $\psi(3686)$ events in the three data samples is $(2712.4\pm14.3)\times10^6$.
 \end{abstract}

 \begin{keyword}
 $\psi(3686)$, inclusive process, hadronic events, BESIII detector
 \end{keyword}


 \begin{multicols}{2}
 \section{Introduction}

 In 2009, 2012, and 2021, the BESIII experiment accumulated the world's largest $\psi(3686)$ data sample produced in electron-positron collisions, thereby providing an excellent platform to precisely study the transitions and decays of the $\psi(3686)$ and its daughter charmonium states, including the $\chi_{cJ}$, $h_c$, and $\eta_c$, and to search for rare decays with physics beyond the Standard Model. The number of $\psi(3686)$ events, $N_{\psi(3686)}$, is a basic input parameter, and its precision has a direct impact on the accuracy of these measurements.

 In this paper, we determine the number of $\psi(3686)$ events by using inclusive $\psi(3686)$ hadronic decays, where the branching fraction of $\psi(3686)\to hadrons$ is known to be $(97.85\pm 0.13)$\%~ \citep{Luth:1975bh,BES:2002psd,ParticleDataGroup:2016lqr}.
 The nonresonant background yield under the $\psi(3686)$ peak is evaluated by analyzing the two off-resonance data samples taken in 2009 and 2021 at a center-of-mass (c.m.) energy $E_{\rm cm} = 3.65$~GeV. The same method of background estimation was successfully used in our previous measurement of the numbers of $\psi(3686)$ events in the data samples collected in 2009 and 2012~\cite{BESIII:2017tvm}.


 \section{BEPCII and BESIII Detector}

 BEPCII~\cite{BESIII:2009fln} is a double-ring $e^+e^-$ collider in the center-of-mass energy range from $2.0 \rm ~to~ 4.95 ~\rm GeV$ which has reached a peak luminosity of $1\times10^{33}~\rm{cm}^{-2}\rm{s}^{-1}$ at $\sqrt{s}$= 3.773 GeV. The cylindrical core of the BESIII detector~\cite{BESIII:2009fln} consists of a helium-based multilayer drift chamber (MDC), a plastic scintillator time-of-flight (TOF) system, and a CsI(Tl) electromagnetic calorimeter (EMC), which are all enclosed in a superconducting solenoid magnet with a field strength of 1.0~T (0.9~T in 2012). The solenoid is supported by an octagonal flux-return yoke with resistive plate counter modules interleaved with steel as a muon identifier. The acceptance for charged particles and photons is 93\% over the $4\pi$ stereo angle. The charged-particle momentum resolution at 1 GeV/$c$ is 0.5\%, and the photon energy resolution at 1 GeV is 2.5\% (5\%) in the barrel (end-caps) of the EMC. The time resolution in the TOF barrel region is 68~ps, while that in the end cap region was 110~ps. The end cap TOF system was upgraded in 2015 using multigap resistive plate chamber technology, providing a time resolution of 60~ps. The MDC encountered the Malter effect due to cathode aging during $\psi(3686)$ data taking in 2012. This effect was suppressed by mixing about 0.2\% water vapor into the MDC operating gas~\cite{Dong:2015kba} and can be well modeled by Monte Carlo (MC) simulation. The other sub-detectors worked well during 2009, 2012, and 2021 operation.

 The BESIII detector is modeled with a MC simulation based on \textsc{geant}4~\cite{Allison:2006ve}. The $\psi(3686)$ produced in the electron-positron collisions are simulated with the generator \textsc{kkmc}~\cite{Jadach:2000ir}, which includes the beam energy spread according to the measurement of BEPCII and the effect of initial state radiation (ISR).  The known decay modes of the $\psi(3686)$ are generated with \textsc{evtgen}~\cite{Ping:2008zz} according to the branching fractions from the Particle Data Group \cite{ParticleDataGroup:2016lqr}, while the remaining unknown decays are simulated using the \textsc{lundcharm} model~\cite{Chen:2000tv}. The MC events are mixed with randomly triggered events (non-physical events from collision) from data to take into account possible effects from beam-related backgrounds, cosmic rays, and electronic noise.


 \section{Event Selection}
 The data collected at the $\psi(3686)$ peak includes several different processes, \(i.e.\), $\psi(3686)$ $\to {\it hadrons}$, $\psi(3686)\to \ell^+\ell^-$ ($\ell=e$, $\mu$ or $\tau$), ISR return to $J/\psi$, and nonresonant background including $e^+e^-\rightarrow{} \gamma^* \rightarrow{} \it hadrons$ ($q\bar q$ ($q=u,d,s$)), $e^+e^-\to \ell^+\ell^-$, and $e^+e^-\rightarrow{}e^+e^- +X$ ($X$= $\it hadrons$, $\ell^+\ell^-$). The data also contains non-collision events, $e.g.$, cosmic rays, beam-associated backgrounds, and electronic noise.

 To separate the candidate events for $\psi(3686)$ $\to$ {$\it hadrons$} from backgrounds, we require that there is at least one  good charged track candidate in each event. The charged tracks are required to be within 10 cm from the Interaction Point (IP) in the $z$ axis, within 1 cm in the perpendicular plane and within a polar angle range of $|\rm cos\theta| < 0.93$ in the MDC, where \(\theta\) is the angle measured relative to the $z$ axis. Photons reconstructed in the EMC barrel region ($|\!\cos\theta|<0.80$) must have a minimum energy of 25 MeV, while those in the end-caps ($0.86<|\cos\theta|<0.92$) must have an energy of at least $50~\rm MeV$. The photons in the polar angle range between the barrel and end-caps are excluded due to the poor resolution. A requirement of the EMC cluster timing [0,~700] ns is applied to suppress electronic noise and energy deposits unrelated to the event.

 The selected hadronic events are classified into three categories according to the multiplicity of good charged tracks ($N_{\rm good}$), $i.e.$, type-I ($N_\text{good}=$1), type-II ($N_\text{good}=2$) and type-III ($N_\text{good}>2$). For the type-III events, no further selection criteria are required.

 \begin{figure*}[htp]
    \centering
      \includegraphics[width=0.45\textwidth]{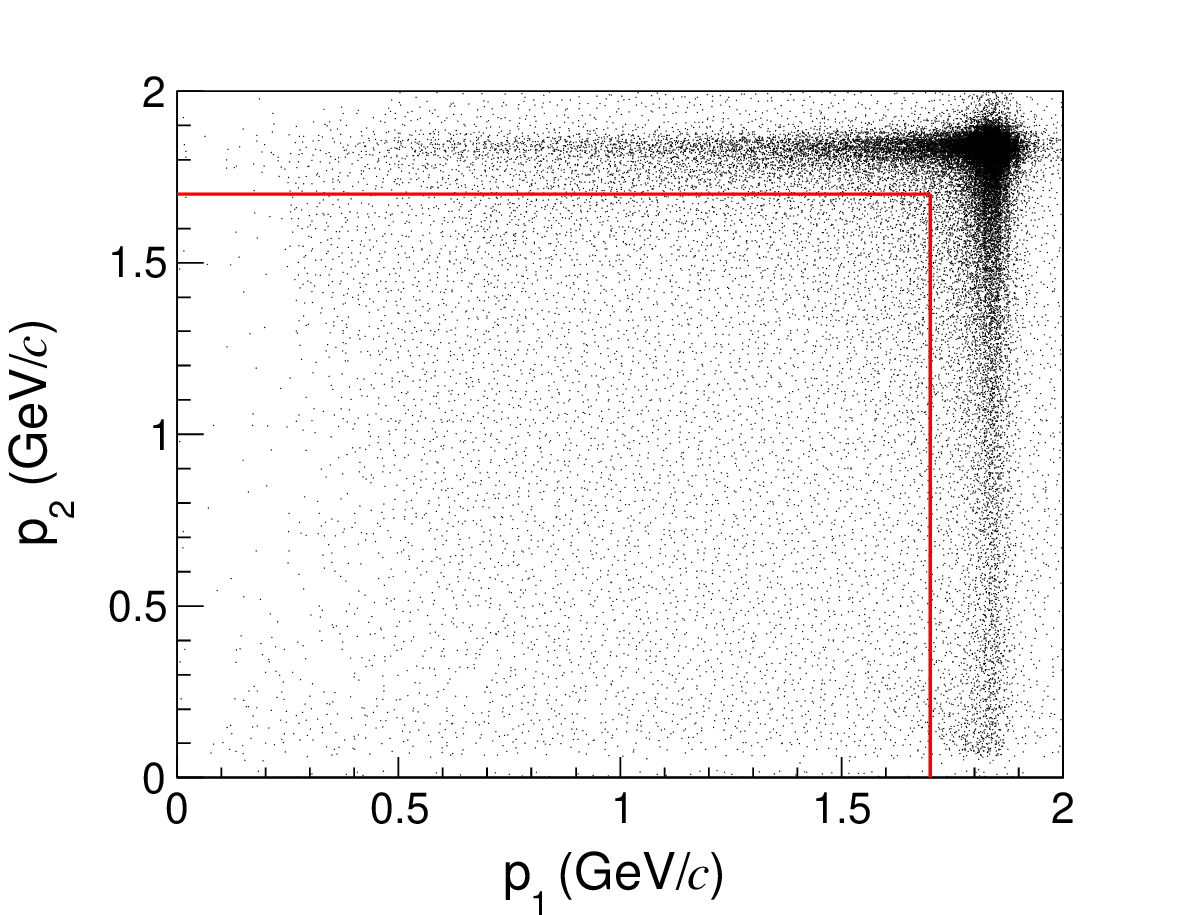}
      \includegraphics[width=0.45\textwidth]{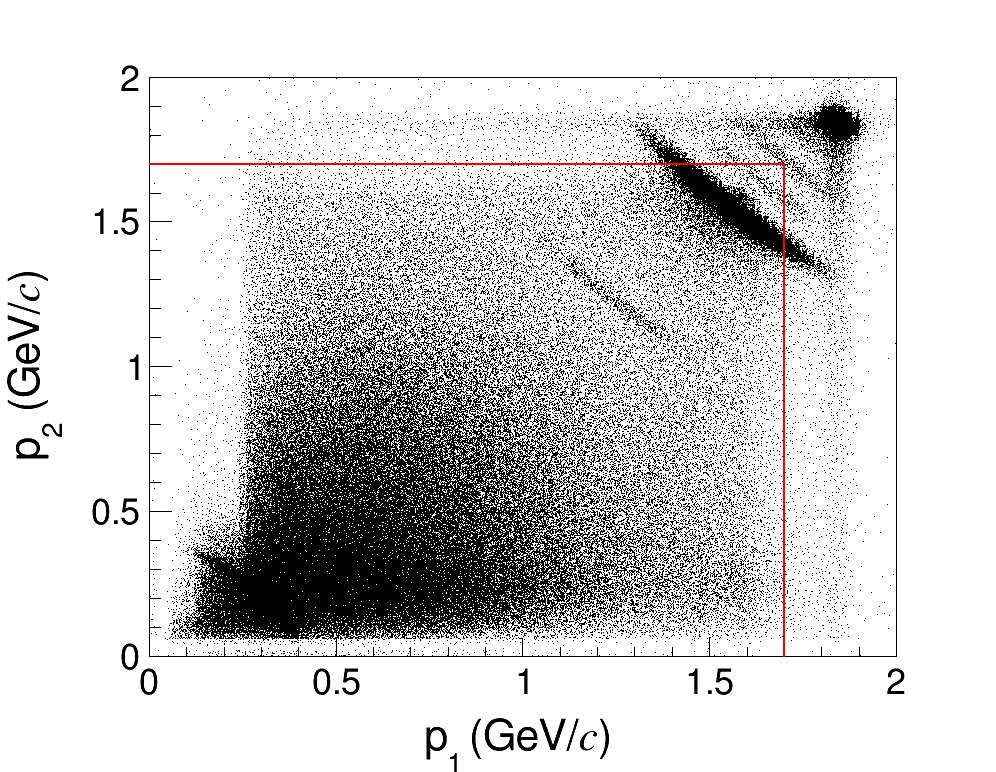}
      \includegraphics[width=0.45\textwidth]{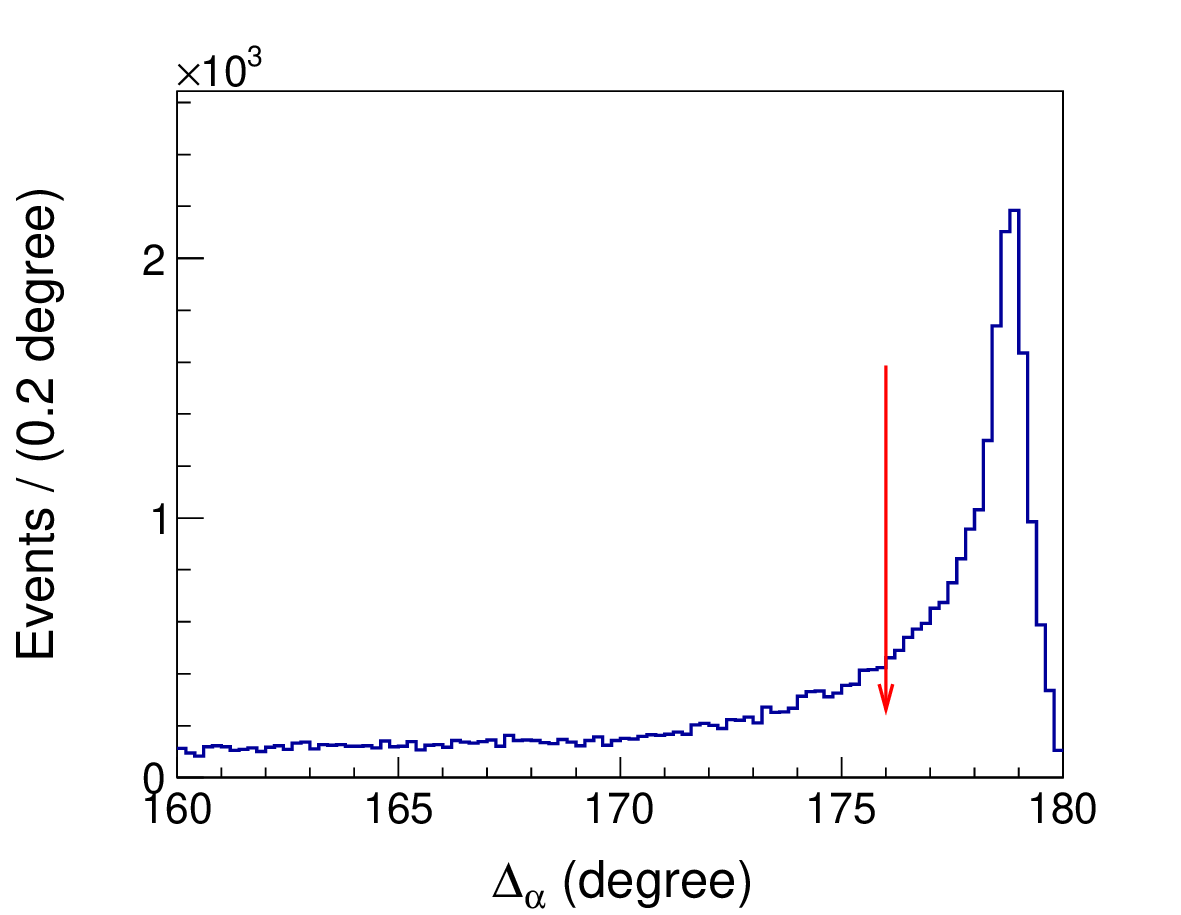}
      \includegraphics[width=0.45\textwidth]{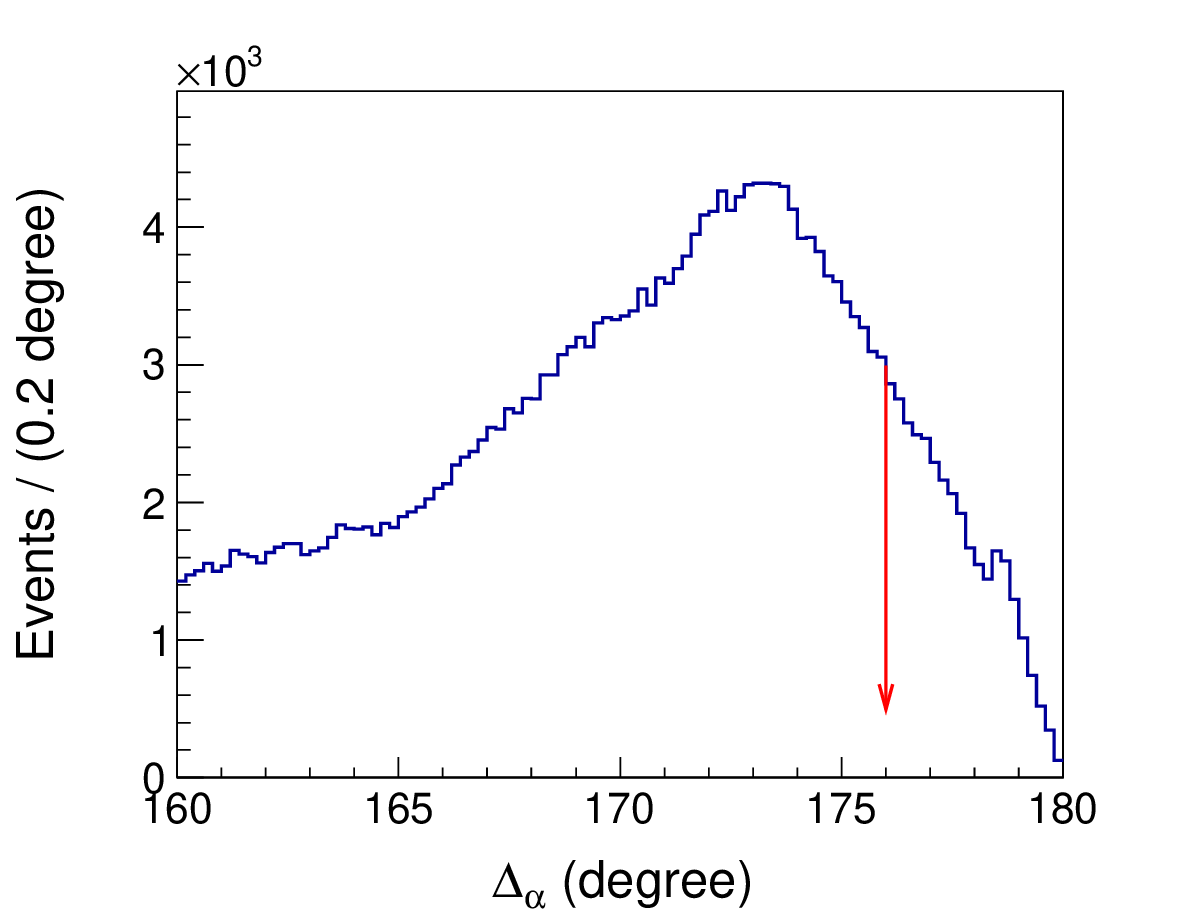}
    \figcaption
    {Distributions of $p_2$ versus $p_1$ (top) and $\Delta_\alpha$ (bottom) for the type-II events from the (left) simulated Bhabha and (right) $\psi(3686)$ inclusive MC samples. The events satisfying $p <$ 1.7 GeV/$c$ and $\Delta_\alpha < 176^{\circ}$ are kept for further analysis. In the top-right plot, the event accumulation in the top-right corner comes from \(\psi(3686)\to\)\( e^{+}e^{-}\), \(\mu^{+}\mu^{-}\), while the different event bands come from \(\psi(3686)\to neutral + J/\psi\), \(J/\psi\to e^{+}e^{-}, \mu^{+}\mu^{-}\), etc., and the event band in the bottom-left comes from \(\psi(3686)\to \pi^{+}\pi^{-}J/\psi\), \(J/\psi\to e^{+}e^{-},\mu^{+}\mu^{-}\) where the lepton pair is missing.
    }
  \label{scatter_figure}
  \end{figure*}
 
  \begin{figure*}[!hbp]
      \centering
      \includegraphics[width=0.45\textwidth]{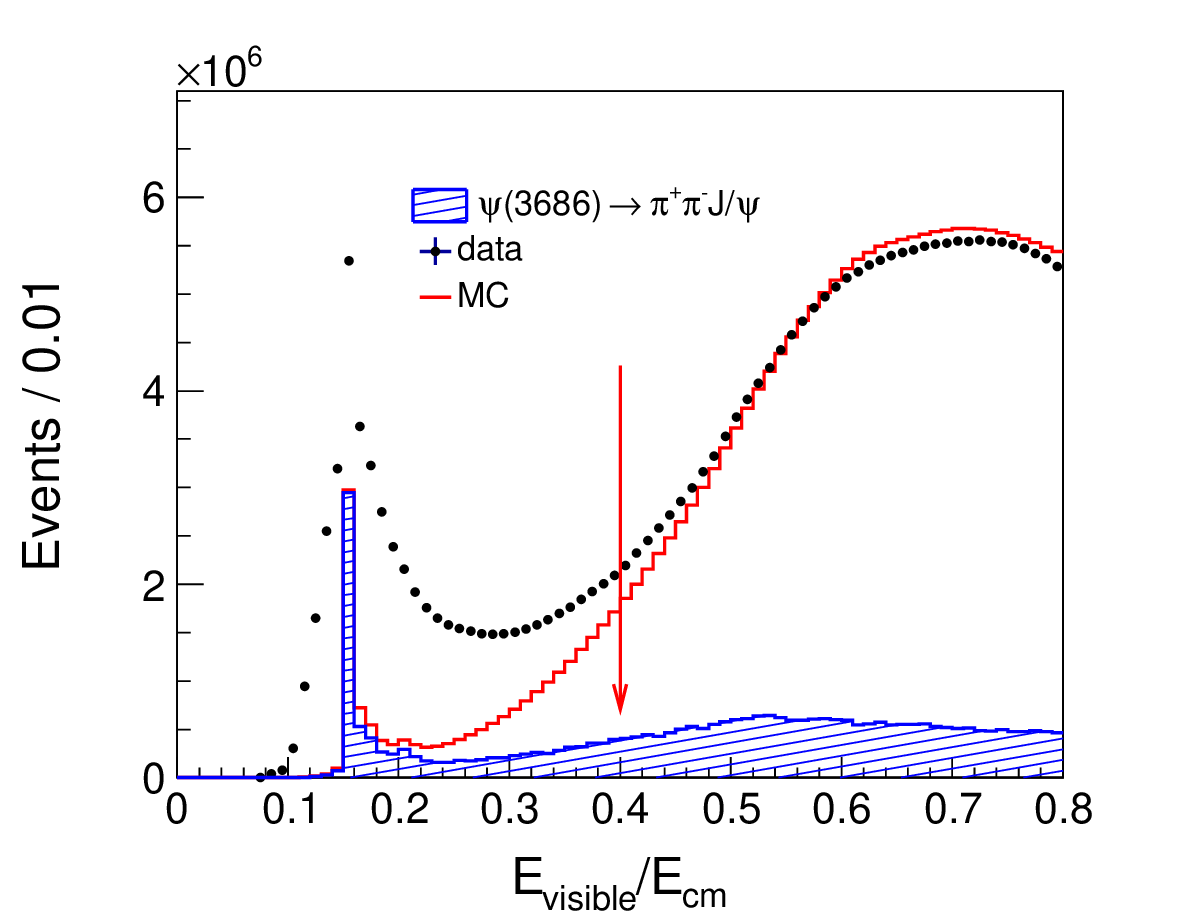}
      \includegraphics[width=0.45\textwidth]{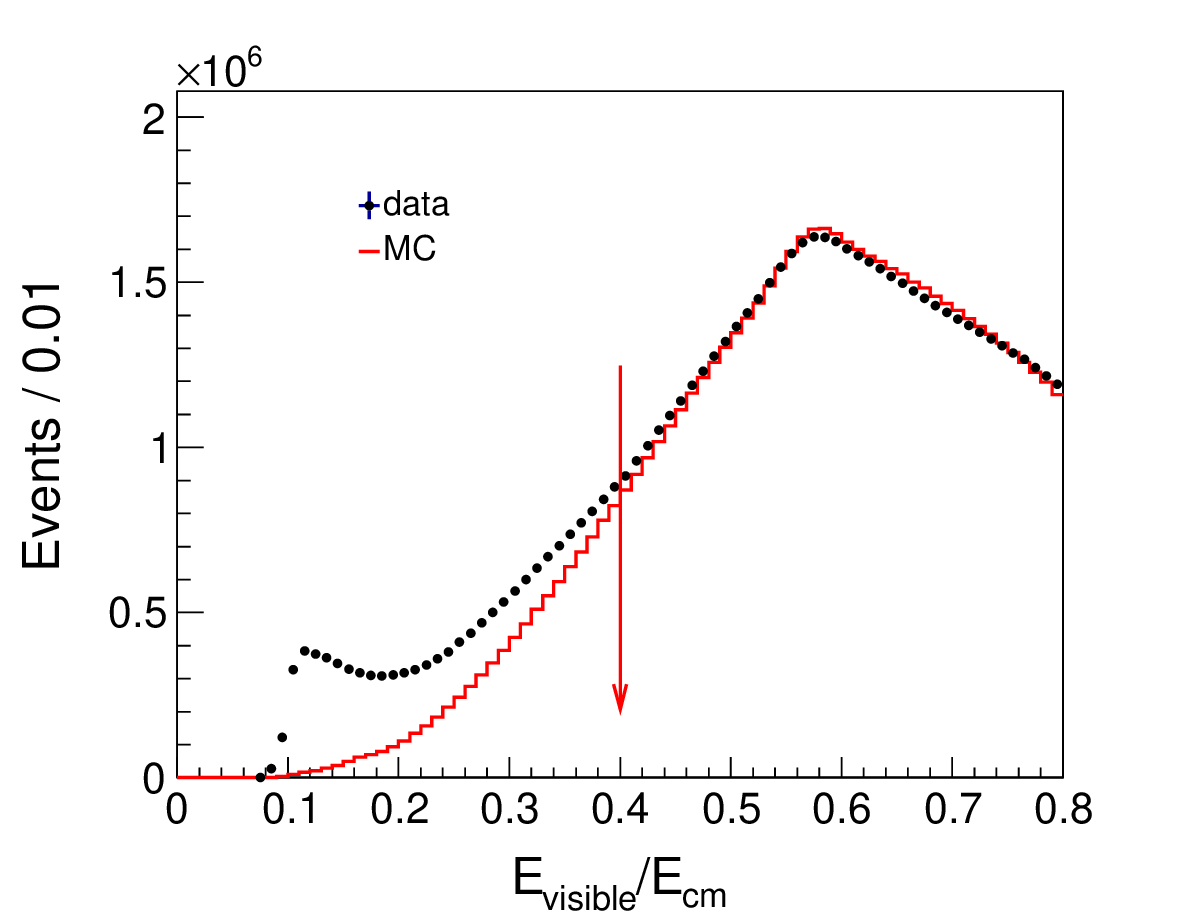}
      \caption
      {Distributions of \(E_{\rm visible}/E_{\rm cm}\) for the type-II (left) and type-I (right) events of the $\psi(3686)$ data and inclusive MC samples. The MC distributions have been scaled to data by using events with \(E_{\rm visible}\)$/$\(E_{\rm cm}>\)~0.4. The events lying above the red arrows are kept for further analysis.
      }
      \label{fig:all_images_ngood_12}
  \end{figure*}

 For the type-II events, the momenta of the two charged tracks ($p_1$ and $p_2$) are required to be less than 1.7~GeV/$c$ and the opening angle between them ($\Delta_\alpha$) is required to be less than $176^{\circ}$ to suppress Bhabha ($e^+e^-\to e^+e^-$) and dimuon ($e^+e^-\to \mu^+\mu^-$) backgrounds. Figure~\ref{scatter_figure} shows the distributions of $p_2$ versus $p_1$ and $\Delta_\alpha$ for the type-II events from the simulated Bhabha and inclusive $\psi(3686)$ MC samples. Furthermore, a requirement of $E_{\rm  visible}/E_{\rm cm}>0.4$ is applied to suppress the low energy background (LEB), which may comprise $e^+e^-\rightarrow{}e^+e^- +X$, double ISR events ($e^+e^- \to \gamma_{\rm ISR} \gamma_{\rm ISR} X$), $etc$. Here, $E_{\rm visible}$ is the visible energy which is the sum of the energy of all the charged tracks calculated with the track momentum assuming the tracks to be pions and all the neutral showers. Figure~\ref{fig:all_images_ngood_12}(left) shows the $E_{\rm  visible}/E_{\rm cm}$ distributions of the type-II events for the $\psi(3686)$ data and inclusive MC samples. The visible excess in inclusive MC at low energy is from \(\psi(3686)\)$\rightarrow{}$\(\pi^{+}\pi^{-}J/\psi\), \(J/\psi\rightarrow{}e^{+}e^{-},~\mu^{+}\mu^{-}\) where the lepton pair is missing. Unless noted, in all plots, the points with error bars are the $\psi(3686)$ data sample collected in 2021, and the histogram is the corresponding inclusive MC sample.

 For the type-I events, at least two photons are required in the event. Compared to those events with high charged track multiplicity, the type-I sample has more background according to the vertex distribution of the charged tracks. Thus, a neutral hadron $\pi^0$ candidate is required to improve the suppression of background events ~\cite{BESIII:2017tvm}, where the $\pi^0$ candidate is reconstructed from a $\gamma\gamma$ pair. Within each event, only the $\gamma\gamma$ candidate with invariant mass closest to the $\pi^0$ nominal mass and satisfying $|M_{\gamma\gamma}-M_{\pi^0}|<$ 0.015 \(\rm GeV\)\(/c^2\) is kept for further analysis. Figure~\ref{mass of pi0} shows the $M_{\gamma\gamma}$ distribution of the selected $\pi^0$ candidates for type-I events. With the above selection criteria, the corresponding $E_{\rm visible}/E_{\rm cm}$ distributions of the candidate events for the $\psi(3686)$ data and inclusive MC samples are shown in Fig.~\ref{fig:all_images_ngood_12}(right). An additional requirement of $E_{\rm visible}/E_{\rm cm}>$ 0.4 is also used to suppress the LEB events.

\begin{figure*}
   \centering
     \includegraphics[width=0.45\textwidth]{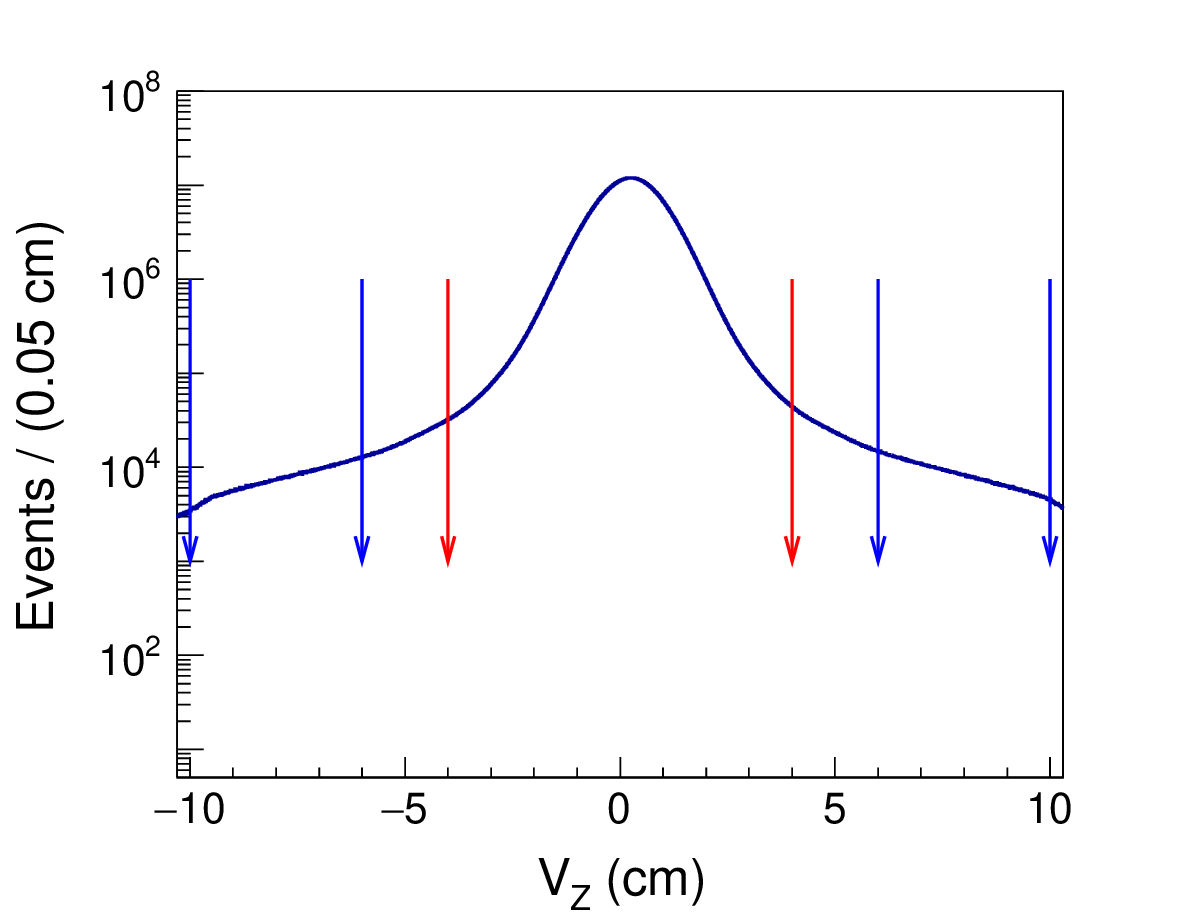}
     \includegraphics[width=0.45\textwidth]{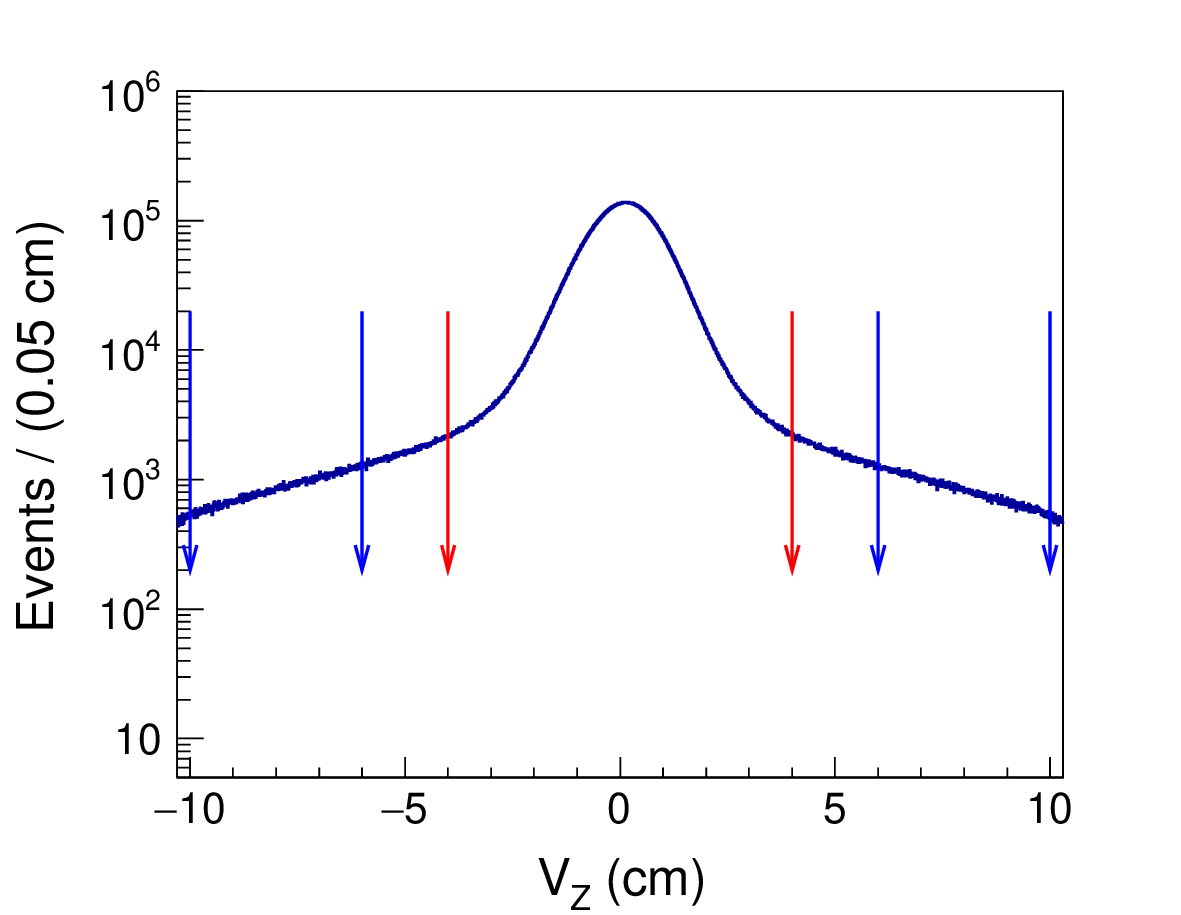}    \captionsetup{justification=raggedright,singlelinecheck=false}
     \figcaption
     {
     The distributions of $V_Z$ for the accepted hadronic events of the (left) \(\psi(3686)\) data and (right) off-resonance data. The region within the pair of red arrows is the signal region, while the regions within the two pairs of blue arrows are the sideband regions.
     }
     \label{fit result vertexfit position}
     \end{figure*}

The signal yield of $e^+e^-\to {\it hadrons}$ is obtained by examining the event vertex distribution $V_Z$. For type-II and type-III events, the $V_Z$ is the event vertex fit position, while for type-I events, the $V_Z$ is the defined one for single track, $i.e$., the distance to IP in the $z$ direction.
Figure \ref{fit result vertexfit position} shows the distributions of $V_Z$ for \(\psi(3686)\) and off-resonance data samples. The region $|V_Z|<4$~cm is regarded as the signal region, and the region $6 <|{V}_Z|<10$~cm is taken as the sideband region. Events in the sideband region are mainly from non-collision background events. The number of observed hadronic events ($N^{\rm obs}$) is determined by
 \begin{equation}
 N^{\rm obs} = N_{\rm signal} - N_{\rm sideband},
 \label{N obs}
 \end{equation}
 where $N_{\rm signal}$ and $N_{\rm sideband}$ are the numbers of events in the signal and sideband regions, respectively.

 \begin{center}
     \includegraphics[width=0.45\textwidth]{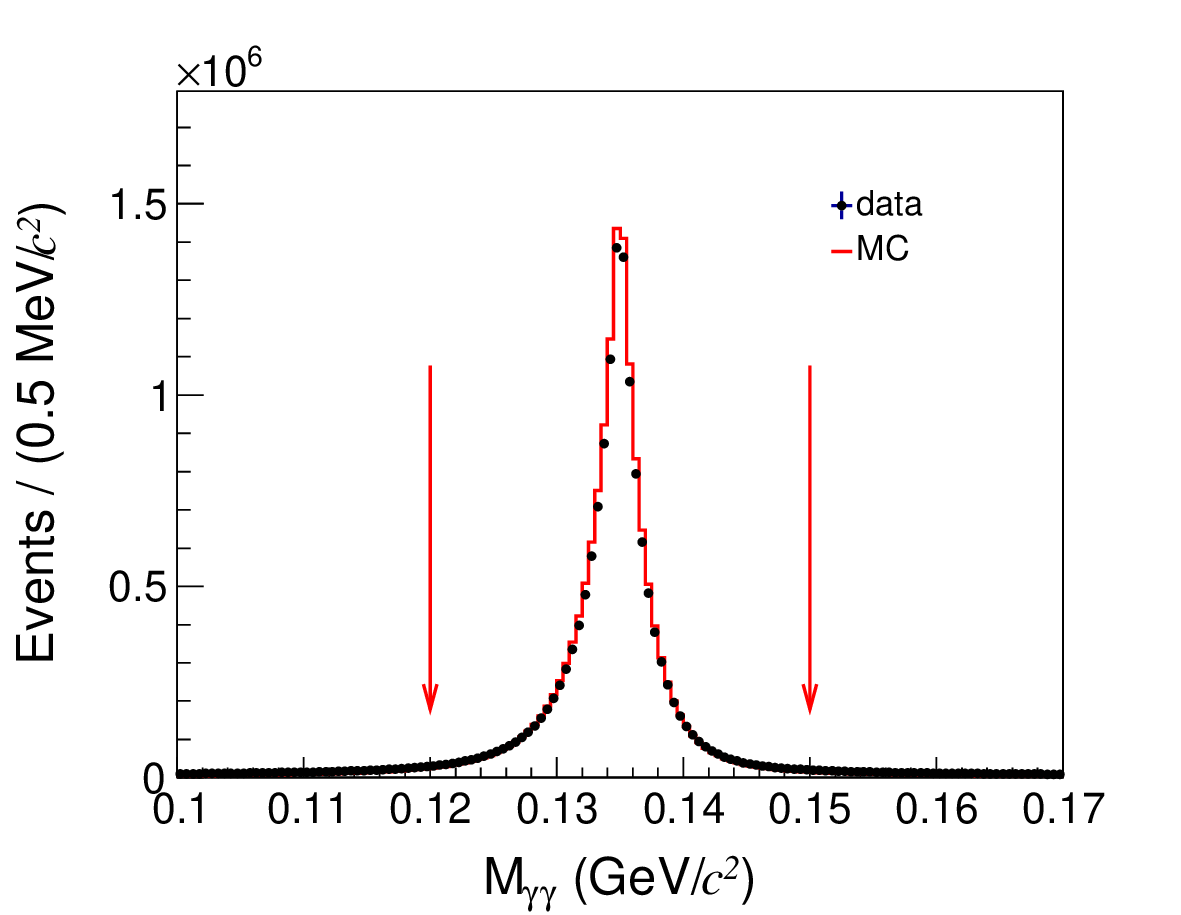}
   \captionsetup{justification=raggedright,singlelinecheck=false}
     \figcaption
     {Distribution of \(M_{\gamma\gamma}\) for the \(\pi^{0}\) candidates from the type-I events. The region within the pair of red arrows is the $\pi^0$ signal window. }
     \label{mass of pi0}
 \end{center}

 \section{Background Analysis}
 Following our previous measurement~\cite{BESIII:2017tvm}, the number of remaining $q\bar q$ background events is estimated using the off-resonance data sample benefiting from the small difference in the c.m. energies, where one expects the $q\bar{q}$ background is comparable so that can be used to estimate the background in the energy of $\psi(3686)$. We apply the same approach to determine the yields of collision events for the off-resonance data samples, then estimate the contribution at the $\psi(3686)$ resonance after scaling by integrated luminosity. Similarly, the contributions from the ISR return to $J/\psi$ are estimated by the same method as above. The connection between two energy points can be expressed by a scaling factor, $f$, determined from the integrated luminosity multiplied by $1/s$ to account for the energy dependence of the cross section. This can be done since the dominant backgrounds come from the Bhabha and dimuon processes at the leading-order contribution~\cite{Balossini:2006wc}.  The scale factor is
 \begin{equation}
 f = \frac{\mathcal{L}_{\psi(3686)}}{\mathcal{L}_{\rm off-res}}\cdot\frac{3.65^2}{3.686^2},
 \label{factor}
 \end{equation}
 where \(\mathcal{L}_{\psi(3686)}\) and \(\mathcal{L}_{ \rm off-res}\) are the integrated luminosities of the \(\psi(3686)\) data and off-resonance data samples, and $3.686^2$ and $3.65^2$ are the corresponding squares of c.m. energies, respectively.

 The integrated luminosities at different energy points are determined using Bhabha events \cite{Ablikim:2013ntc} with the following selection criteria. The number of charged tracks is required to be equal to two with net charge zero. Each track must have energy deposited in the EMC between 1.0 GeV and 2.5 GeV and a momentum less than $0.5\times E_{\rm cm} + 0.3$ GeV.
 Furthermore, the sum of the momenta of the positron and electron must be greater than 0.9\(\times E_{\rm cm}\). The cosine of the polar angle ($\theta$) for each track is required to be within \(\left| \rm cos\theta \right| < 0.8 \) and their $\phi$ angles must satisfy \( 5^{\circ} <\left| \left| \phi_{1} - \phi_{2} \right| - 180^{\circ}  \right| < 40^{\circ} \). The luminosities of the \(\psi(3686)\) and off-resonance data samples taken in 2021 are 3208.5 pb$^{-1}$ and 401.0 pb$^{-1}$, with uncertainties of about 1\%, respectively.

 To test if the interference between Bhabha events and  $\psi(3686)\to e^+e^-$ events affects the luminosity measurement, we examine the integrated luminosities of the $\psi(3686)$ data samples using $e^+e^-\to\gamma\gamma$ events with the following selection criteria. The number of showers is required to be greater than or equal to two with no candidate charged tracks. Each shower must have deposited energy in the EMC between 1.0 GeV and $2.5~\rm GeV$. The cosine of the polar angle ($\theta$) for each shower is required to be within \(\left| \rm cos\theta \right| < 0.8 \).
 The two most energetic showers are required to be back to back (\( ||\theta_1 - 90^{\circ}| - |\theta_2 - 90^{\circ}|| < 10^{\circ}\)) and with $\phi$ angles \( \left| \left| \phi_{1} - \phi_{2} \right| - 180^{\circ}  \right| < 2^{\circ} \). The difference of the measured luminosities is less than 0.1\%.

 The integrated luminosities of the two off-resonance data samples collected at $E_{\rm cm}=3.65$ GeV in 2009 and 2021 are 44.5 and 401.0 $\rm pb^{-1}$, respectively. The former one is used to estimate the continuum  contribution of the 2009 $\psi(3686)$ data sample, and the latter one is used to estimate the continuum  contribution of the 2012 and 2021 $\psi(3686)$ data samples. The integrated luminosities of the 2009, 2012, and 2021 $\psi(3686)$ data samples are 161.6, 506.9, and 3208.5 pb$^{-1}$, respectively, with  scaling factors $f$ of 3.56, 1.24, and 7.85, respectively. The systematic uncertainties of the luminosities for the two c.m. energies almost cancel when calculating the scaling factors due to the small energy difference.

 The cross sections for $e^{+}e^{-}\rightarrow{}\tau^{+}\tau^{-}$ are calculated to be 1.84 and 2.14~nb at $E_{\rm cm}=3.65\;\text{GeV}$ and 3.686 GeV, respectively. Since the above energy points are close to the $\tau^{+}\tau^{-}$ mass threshold, the production cross section does not follow a $1/s$ distribution. Thus, only part of the $e^{+}e^{-}\rightarrow{}\tau^{+}\tau^{-}$ background events is included in the off-resonance data samples. To subtract the full background from $e^{+}e^{-}\rightarrow{}\tau^{+}\tau^{-}$, we estimate the remaining contribution, $N^{\rm uncanceled}_{\tau^+\tau^-}$, using the detection efficiency from the MC simulation, the cross section difference at the two c.m. energy points, and the luminosity at the $\psi(3686)$ peak. The estimated residual $e^+e^-\to \tau^+\tau^-$ background yields are shown in Table~\ref{number result}.

 The cross section from the ISR return to $J/\psi$ is also found to slightly violate the $1/s$ distribution. However, the corrected cross section difference for this process at the $\psi(3686)$ peak is about $0.1 ~\rm nb$, which is negligible if compared to the total observed cross section of $\psi(3686)\rightarrow{}\it hadrons$, $\sim$ 700 $\rm nb$.

 A small fraction of $\psi(3686)\to \ell^+\ell^-$ events  survives the event selection. Since their effect has been considered in the detection efficiency, no further subtraction is made.

\begin{figure*}
   \centering
     \includegraphics[width=0.45\textwidth]{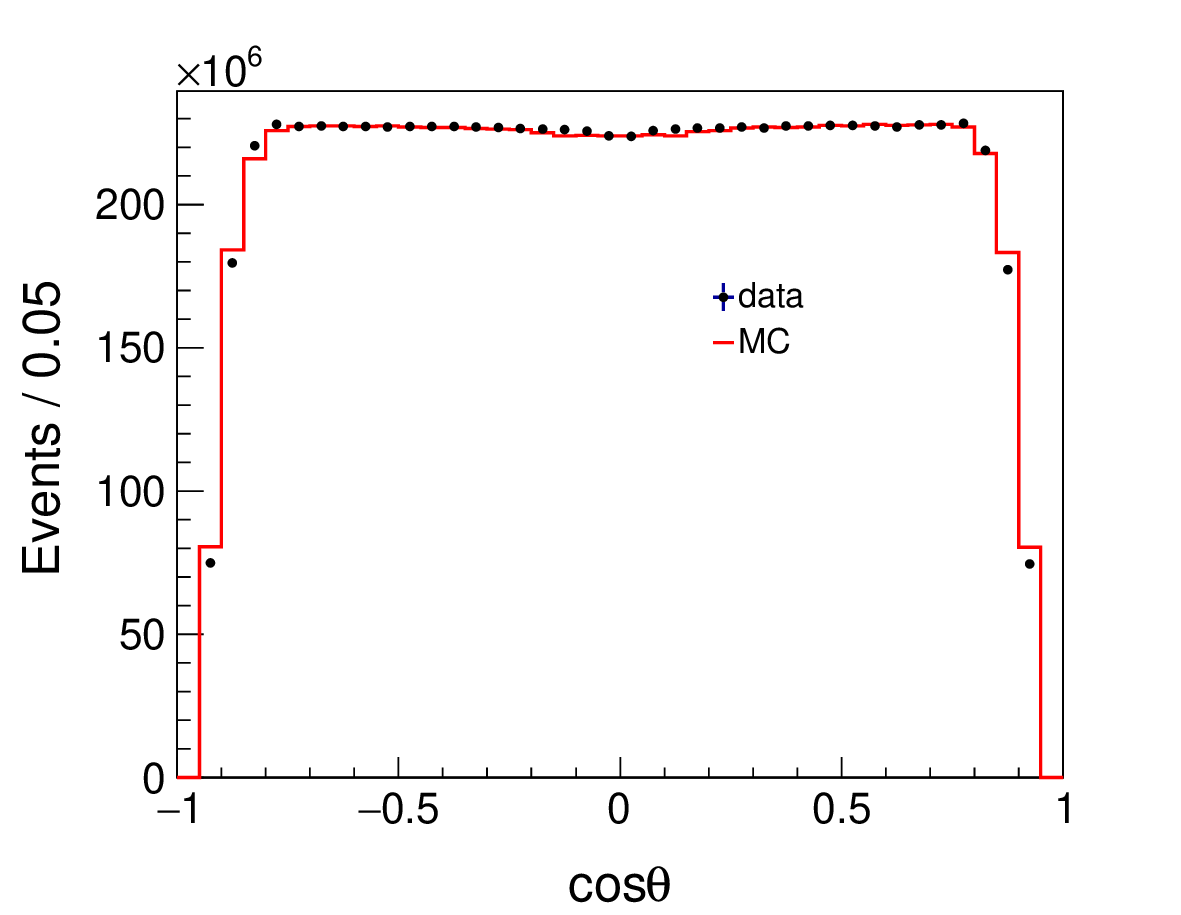}
     \includegraphics[width=0.45\textwidth]{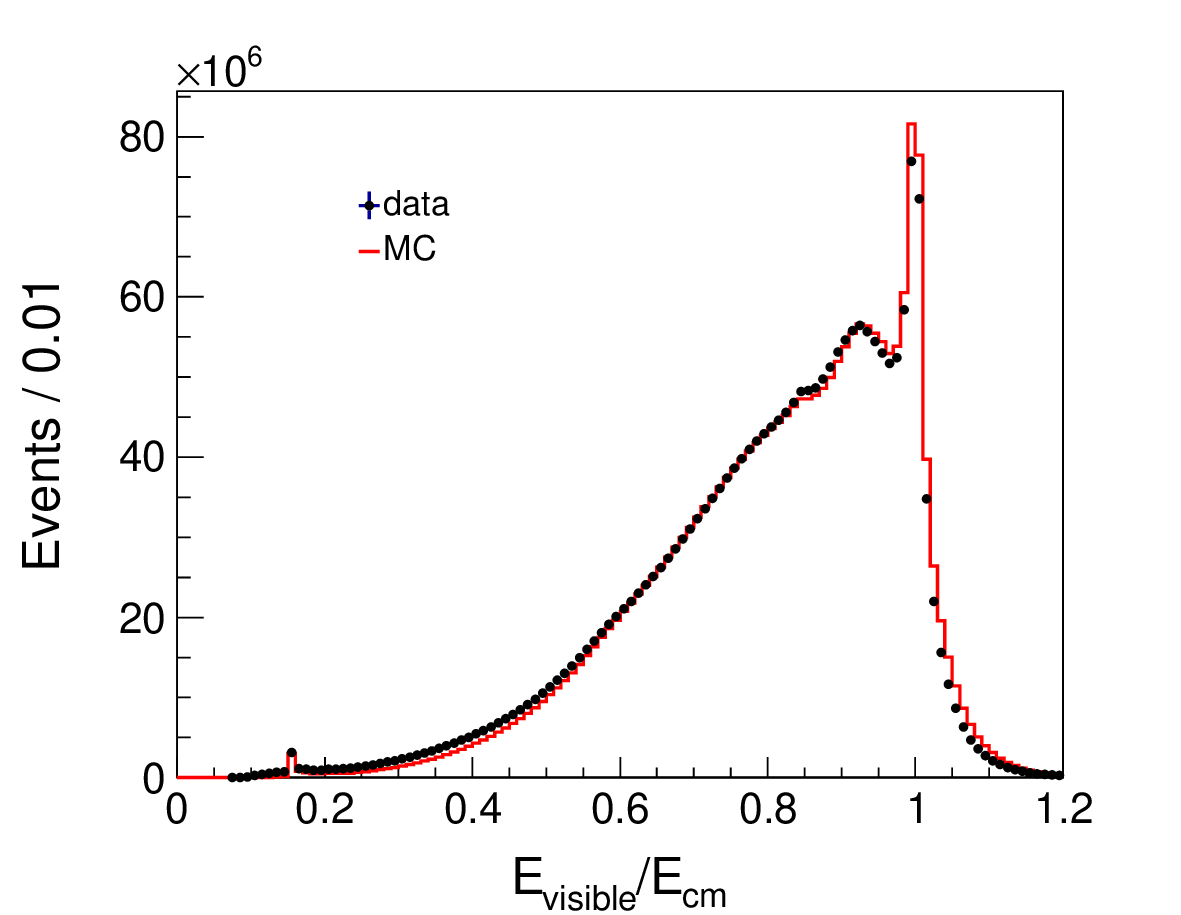}
     \includegraphics[width=0.45\textwidth]{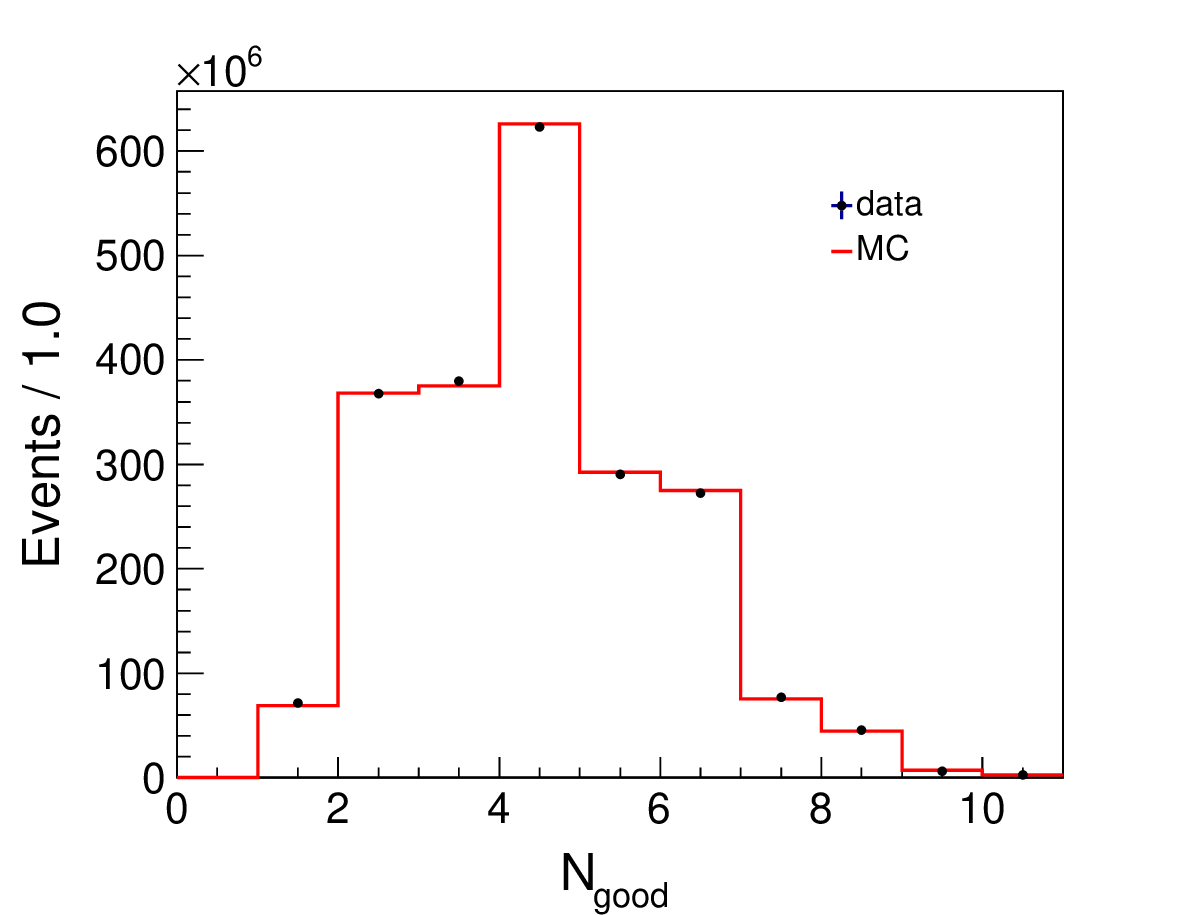}
     \includegraphics[width=0.45\textwidth]{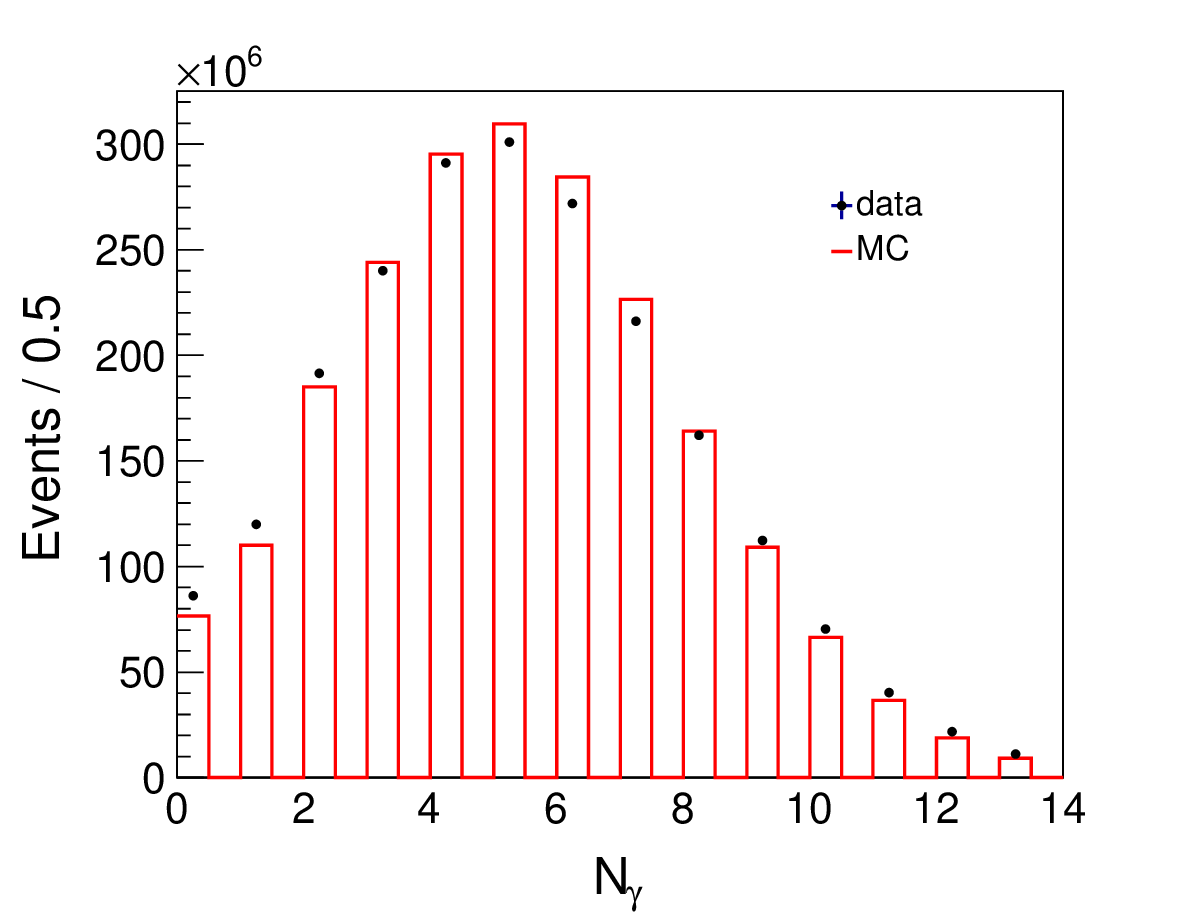}

     \captionsetup{justification=raggedright,singlelinecheck=false}
   \caption{Comparisons of the distributions of (top-left) \(\rm cos\theta\), (top-right) \(E_{\rm visible}/E_{\rm cm}\), (bottom-left) $N_{\rm good}$, and (bottom-right) $N_{\gamma}$ between $\psi(3686)$ data and inclusive MC samples after background subtraction.}
   \label{plot after backgruond subtraction}
 \end{figure*}

  Figure~\ref{plot after backgruond subtraction} shows the comparisons of the distributions of $\cos\theta$, $E_{\rm visible}/E_{\rm cm}$, $N_{\rm good}$, and photon multiplicity ($N_{\gamma}$) after background subtraction between data and MC simulation, and a reasonable data-MC agreement is observed.
 Table~\ref{number result} summarizes the numbers of the observed hadronic events for different $N_{\rm good}$ requirements of $\psi(3686)$ data ($N^{\rm obs}_{\psi(3686)}$) and off-resonance data ($N^{\rm obs}_{\rm off-res}$). The detection efficiencies of $\psi(3686)\rightarrow{}{\it hadrons}$ are determined with 2.3 billion $\psi(3686)$ inclusive MC events, where the branching fraction of $\psi(3686)\rightarrow{} {\it hadrons}$ is included in the efficiency.

 \begin{table*}[htbp]
     \captionsetup{justification=raggedright,singlelinecheck=false}
   \caption{Number of hadronic events  \(N^{\rm obs}_{\psi(3686)}\) in the \(\psi(3686)\) data, separately for different requirements on the number of good tracks $N_{\rm good}$,	
   where \(N^{\rm obs}_{\psi(3686)}\) is the number of hadronic events observed in the \(\psi(3686)\) data, $f$ is the scaling factor, \(N^{\rm obs}_{\rm off-res}\) is the number of  hadronic events observed in the off-resonance data, \(N^{\rm uncanceled}_{\tau^{+}\tau^{-}}\) is the number of remaining \(e^{+}e^{-}\rightarrow{}\tau^{+}\tau^{-}\) events after subtracting the normalized off-resonance data, \(\epsilon\) is the detection efficiency, and \(N_{\psi(3686)}\) is the determined number of \(\psi(3686)\) events. The statistical uncertainties are expected to be negligible.}
   \centering
     \renewcommand*{\arraystretch}{1.5}
     \setlength{\tabcolsep}{2.5mm}{
 \begin{tabular}{c c c c | c c c | c c c}
  \hline
  Multiplicity &                     \multicolumn{3}{c}{$N_{\rm good}\geq1$}&       \multicolumn{3}{c}{$N_{\rm good}\geq2$}   &\multicolumn{3}{c}{$N_{\rm good}\geq 3$}\\
  \hline
   Year        &                         2009  &     2012&      2021&                2009&       2012&        2021&                  2009&    2012&     2021\\
 $N^{\rm obs}_{\psi(3686)} (10^6)$ &                 107.98&    345.14&     2246.93&            104.77&        333.79&      2172.16&             83.36&      264.57&  1722.56\\
 $f$  &                                   3.56 & 1.24&   7.85&                         3.56 & 1.24&   7.85&                          3.56 & 1.24&   7.85\\
 $N^{\rm obs}_{\rm off-res} (10^6)$ &                   2.05  &    18.18&     18.18&               1.99  &       17.65&        17.65&                0.75 &     6.51&    6.51\\
 $N^{\rm uncanceled}_{\tau^{+}\tau^{-}} (10^6)$ &    0.04  &    0.13&      0.80&                0.04  &      0.12&          0.76&                0.01 &       0.03 &    0.22 \\

 $\epsilon$ (\%)&                        93.21 &    92.83&     92.86&              90.42 &         89.65&       89.86&                74.69&     73.39&    73.85\\  \hline
 $N_{\psi(3686)}~(10^6)$&                      107.96&    347.36&     2265.08&            107.99&       347.75&         2262.32&              108.03&    349.46&    2262.94\\
 \hline
 \end{tabular}}
     \label{number result}
 \end{table*}

 \section{Numerical Results}
 With the numbers listed in Table \ref{number result}, we determine the number of $\psi(3686)$ events using
 \begin{equation}
 N_{\psi(3686)} = \frac{N^{\rm obs}_{\psi(3686)} - f\cdot N^{\rm obs}_{\rm off-res} - N^{\rm uncanceled}_{\tau^{+}\tau^{-}}}{\epsilon}.
 \end{equation}
 The obtained numerical results for $N_{\psi(3686)}$ with different $N_{\rm good}$ requirements are slightly different with each other, mainly due to the imperfect simulation of the charged track multiplicity. To obtain a more accurate $N_{\psi(3686)}$, an unfolding method is employed based on an efficiency matrix determined from the $\psi(3686)$ inclusive MC sample. In practice, there are even numbers of charged tracks generated in an event due to charge conservation, while any number of charged tracks can be observed due to the reconstruction efficiency and backgrounds. Therefore, the true multiplicities of charged tracks of the data sample is estimated from the observed multiplicities of charged tracks and the efficiency matrix by minimizing the $\chi^2$, defined as
 \begin{equation}
 \chi^{2} = \sum_{i = 0}^{10}\frac{(N^{\rm obs}_{i} - \sum_{j = 0}^{10}\epsilon_{ij}\cdot N_{j})^{2}}{N^{\rm obs}_{i}} ,
 \label{efficiency matrix}
 \end{equation}
 where the values $N^{\rm obs}_i~(i=0,~1,~2,\cdots)$ are the observed multiplicities of charged tracks in the data sample corresponding to the distribution in Fig.~\ref{plot after backgruond subtraction}~(bottom-left, the points with error bars),  the matrix elements $\epsilon_{ij}$ represent the probability to observe $i$ charged tracks for an event with $j$ actual charged tracks, and the values $N_j~(j=0,~2,~4,\cdots)$ are the true multiplicities of charged tracks in the data sample. They are free parameters in the fit. For simplicity, the events with ten or more
 charged tracks are combined in the number $N_{10}$. The $N_{\psi(3686)}$ is calculated by summing over all the obtained $N_{j}$. The results are $107.7\times 10^6$, $345.4\times 10^6$ and $2259.3\times 10^6$ for
 the 2009, 2012 and 2021 data samples, respectively.

 \section{Systematic Uncertainties}
 The systematic uncertainties in the $N_{\psi(3686)}$ measurement from different sources are described below and listed in Table~\ref{systematic uncertainty}.

 \subsection{Polar angle of charged tracks}
 The polar angles of charged tracks are required to satisfy $|\cos\theta|<0.93$. To estimate the relevant systematic uncertainty, we redo the measurement with an alternative requirement of  $|\cos\theta|<$0.8. The difference in the measured number of $\psi(3686)$ events is taken as the systematic uncertainty.

 \subsection{Tracking efficiency }
 The systematic uncertainties due to the tracking efficiency for both the 2009 and 2012 data samples have been found to be negligible based on various studies~\cite{BESIII:2017tvm}.
 Therefore, the associated systematic uncertainty for the 2021 data sample is also ignored.

 \subsection{Momentum and opening angle}
 To estimate the systematic uncertainty due to the requirement on charged track momentum for the type-II events, we vary the nominal requirement from $p<1.7$ GeV/$c$ to $p<1.55$ GeV/$c$, and the opening angle between two charged tracks from $\theta<176^{\circ}$ to $\theta<160^{\circ}$. The change in $N_{\psi(3686)}$ is taken as the corresponding systematic uncertainty.

 \subsection{LEB contamination}
 In the nominal measurement, the $E_{\rm visible}/E_{\rm cm}<0.4$ requirement is used to suppress the LEB background events for the type-I and type-II events. The systematic uncertainty due to this requirement is assigned with alternative requirements of  $E_{\rm visiable}/E_{\rm cm}<0.35$ and $E_{\rm visiable}/E_{\rm cm}<0.45$. The larger change in $N_{\psi(3686)}$ is assigned as the systematic uncertainty.

 \subsection{Extraction method of \texorpdfstring{\(N^{\rm obs}\)}{Lg}}
 The nominal measurement is performed by counting the events in the $V_Z$ distributions in  Fig.~\ref{fit result vertexfit position}. To examine the systematic uncertainty associated with the counting method, we use an alternative method by fitting the $V_Z$ distributions with three linearly added Gaussian functions to model the signal shape and a second order polynomial function to describe the non-collision background. The difference in the determined $N_{\psi(3686)}$ values between these two methods is taken as the systematic uncertainty.

 \subsection{Vertex requirement}
 To estimate the systematic uncertainties due to the vertex requirement, we examine the number of $\psi(3686)$ events after varying the  nominal vertex requirements of $V_r<1\;\text{cm}$ to $V_r<2\;\text{cm}$, and from $|V_z|<10$~cm to $|V_z|<15$~cm. The change in the measured number of $\psi(3686)$ events is taken as the systematic uncertainty.

 \subsection{Scaling factor}
 The systematic uncertainty due to the scaling factor $f$ is estimated with the alternatively measured luminosities  with \(e^{+}e^{-}\rightarrow{}\gamma\gamma\) events. The change of the re-measured $N_{\psi(3686)}$ is found to be negligible and this systematic uncertainty is therefore neglected.

 \subsection{\texorpdfstring{\(\pi^{0}\)}{Lg} mass requirement}
 A requirement of $|M_{\gamma\gamma}-M_{\pi^0}|<0.015\;\text{GeV}/c^2$ has been imposed on the type-I events to suppress background. Its effect on the measured number of $\psi(3686)$ events is studied with an alternative requirement of  $|M_{\gamma\gamma}-M_{\pi^0}|<0.025\;\text{GeV}/c^2$. The change in $N_{\psi(3686)}$ is taken as the systematic uncertainty.

 \begin{figure*}
   \centering
     \includegraphics[width=0.45\textwidth]{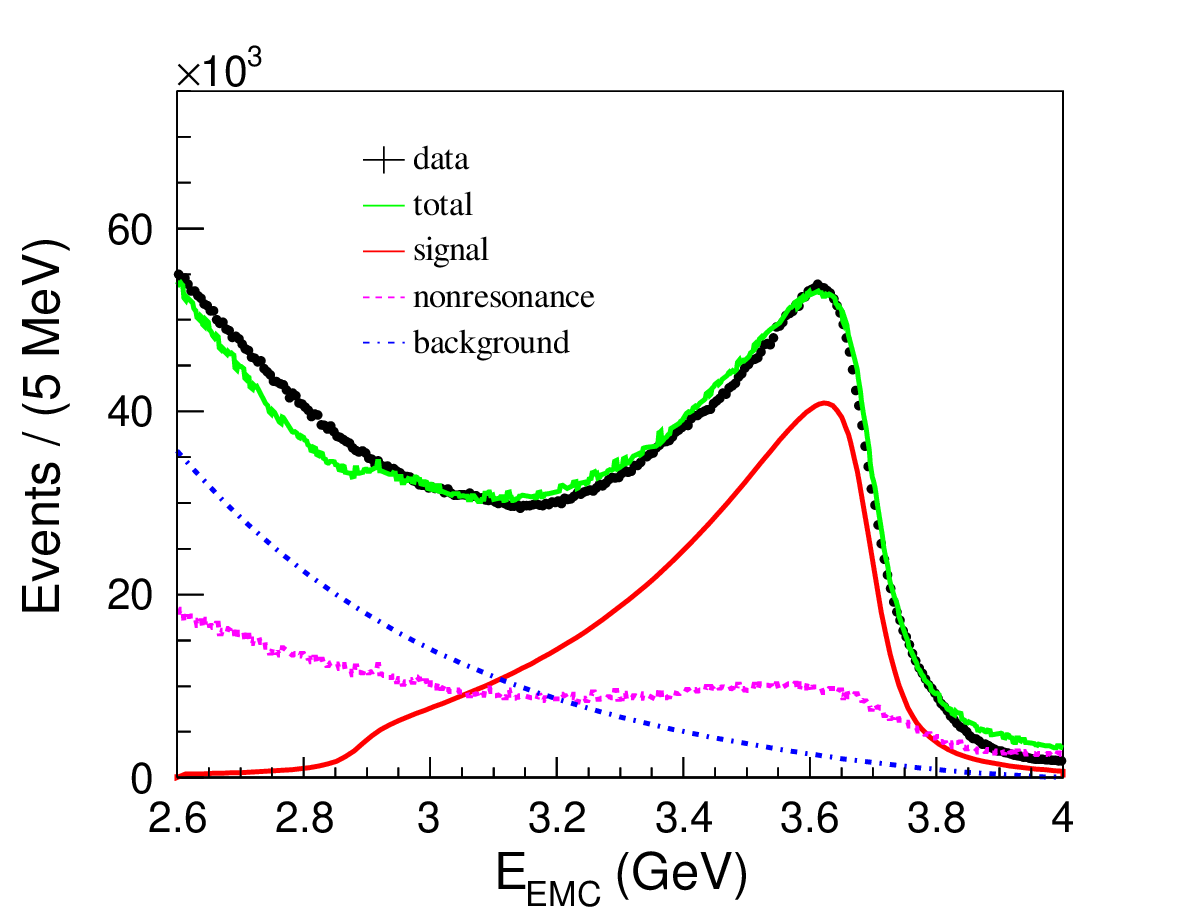}
     \includegraphics[width=0.45\textwidth]{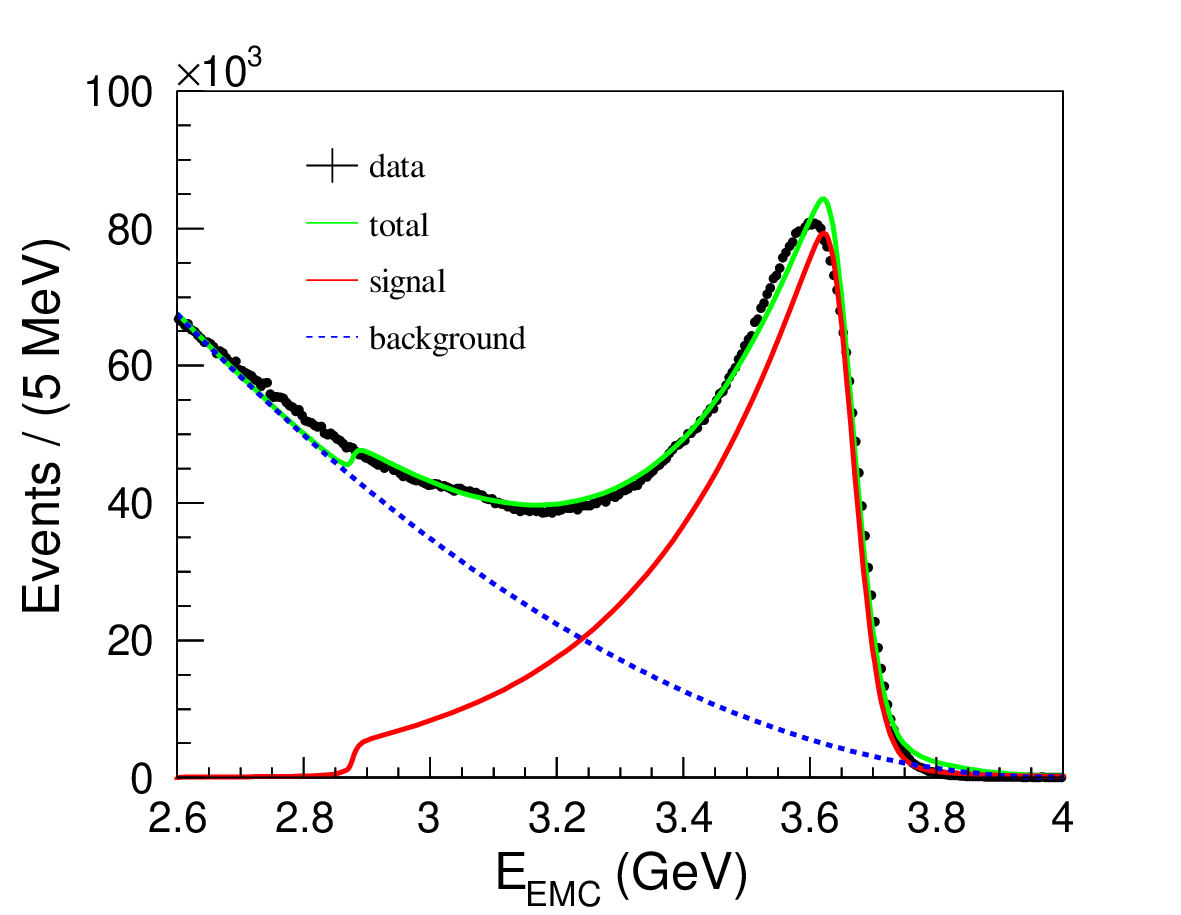}
     \captionsetup{justification=raggedright,singlelinecheck=false}
     \caption{Distributions of $E_{\rm EMC}$ for the $N_{\rm good}=0$ hadronic events from the \(\psi(3686)\) data (left) and inclusive MC (right) samples. The black points with error bar are data. The red lines are the signal shapes of neutral \(\psi(3686)\) decays, the blue lines are the background shapes from \(\psi(3686)\) decays, the pink line is the total background shape from nonresonant processes, and the green lines are the final fit curves. }
     \label{ngood = 0 events}
 \end{figure*}
 \begin{table*}[htbp]
 \captionsetup{justification=centering,singlelinecheck=false}
 \centering
 \caption{Relative systematic uncertainties (\%) in the determination of the number of $\psi(3686)$ events.   }
 \renewcommand*{\arraystretch}{1.3}
 \setlength{\tabcolsep}{7.5mm}{
 \begin{tabular}{c c c c}
 \hline
 Source&                                            2009&               2012&               2021\\         \hline
 Polar angle of charged track&                                  0.25       &             0.20       &         0.22        \\
 Tracking efficiency&                                     negligible &             negligible &       negligible    \\
 Momentum and opening angle&                   0.20       &             0.26       &          0.26       \\
 LEB contamination&                            0.02       &           0.04         &          0.12       \\
 Extraction method of $N^{\rm obs}$ &                  0.16       &           0.16         &          0.03       \\
 Vertex requirement &                          0.13       &           0.08         &         0.07        \\
 Scaling factor ($f$) &                        negligible &           negligible   &          negligible \\
 $\pi^0$ mass requirement &                    negligible &            0.01        &          0.05       \\
 Missing $N_{\rm good} = 0$~ hadronic events&                        0.38      &            0.31        &          0.11       \\
 Charged track multiplicity&                   0.24       &             0.56       &         0.26        \\
 MC modeling&                                  negligible &            negligible  &          negligible \\
 Trigger efficiency&                                      negligible &             negligible &          negligible \\
 $\mathcal{B}(\psi(3686)\rightarrow{}{\it hadrons})$&        0.13   &             0.13       &          0.13\\\hline
 Total&                                          0.60     &            0.75        &         0.49\\\hline
 \end{tabular}}
 \label{systematic uncertainty}
 \end{table*}

 \subsection{Missing \texorpdfstring{\(N_{\rm good}\)}{Lg} = 0 hadronic events}
 We do not consider the $N_{\rm good}=0$ hadronic events in the nominal measurement. The topological analysis for the $\psi(3686)$ inclusive MC samples shows that most of these events come from well-known decay channels, such as $\psi(3686)\rightarrow{} X+ J/\psi$~(where X denotes $\eta$, $\pi^0$, $\pi^0\pi^0$, $\gamma\gamma$, etc.) and $\psi(3686)\rightarrow{}e^{+}e^{-}$, $\mu^+\mu^-$. The fraction of $N_{\rm good} = 0$ hadronic events is $\sim$2.0\%, and the pure neutral channels contribute about 1.0\%.

 We examine the effect of the $N_{\rm good}=0$ hadronic events as follows. A sample is selected using the requirements $N_{\rm good}=0$ and $N_{\gamma}>3$, where the good charged tracks and showers are selected with the same criteria mentioned above.  The $N_{\gamma}>3$ requirement is used to suppress the \(e^{+}e^{-}\rightarrow{}\gamma\gamma\) and beam-associated background events. Figure~\ref{ngood = 0 events} shows the distributions of the total energy in the EMC, $E_\text{EMC}$, for the different data sets and inclusive MC samples. The events concentrated around the c.m. energy are mainly from the pure neutral hadronic candidates. The number of signal events is determined by a fit to the $E_\text{EMC}$ distribution.  In this fit, the signal is described by a Breit-Wigner function convolved with Crystal Ball function, the nonresonant background in the $\psi(3686)$ data sample is described by the shape of off-resonance data sample after luminosity normalization, and the other backgrounds are described by a polynomial function. For the 2021 data sample, the difference in the number of pure neutral hadronic events between the $\psi(3686)$ data and inclusive MC samples is 11\%.  Since the fraction of the pure neutral hadronic events is about 1\% of the total selected
 hadronic events, the systematic uncertainty due to the missing $N_{\rm good} = 0$ hadronic events must be less than $11\%\times1\% = 0.11\%$ for the 2021 data sample. With the same
 method, the systematic uncertainties for the 2009 and 2012 data samples are assigned as 0.38\% and 0.31\%, respectively.

 \subsection{Charged track multiplicity}
 To estimate the systematic uncertainty arising from the charged track multiplicity, we compare the directly calculated result as shown in Table~\ref{number result} and that obtained with the unfolding method after including the $N_{\rm  good}\leq 1$ events. The differences in the numbers of $\psi(3686)$ events for the 2009, 2012 and 2021 data samples, which are 0.24\%, 0.56\%, and 0.26\%, respectively, are taken as individual systematic uncertainties.

 \subsection{MC modeling}
 The systematic uncertainties due to MC modeling include the input branching fractions and the angular distributions of the known and unknown decay modes in the $\psi(3686)$ inclusive MC sample. These uncertainties have been covered by those of the charged track multiplicity and the missing $N_{\rm good} = 0$ events. Hence no systematic uncertainty is assigned for the MC modeling.

 \subsection{Trigger efficiency}
 The trigger efficiencies for BESIII data were studied in 2010~\cite{Berger:2010my} and 2021~\cite{BESIII:2020zpl}. The trigger efficiency for the $N_\text{good}\geq2$ (type-II and type-III) events is found to be close to 100.0\%,
while it is 98.7\% for the type-I events~\cite{Berger:2010my}.
 Since the fraction of the type-I events is only about 3\% of the total selected hadronic events, the associated systematic uncertainty is negligible. The neutral channel trigger has been added since 2012, and the trigger efficiency for the type-I events is expected to be higher than before. Therefore, the systematic uncertainty associated with the trigger efficiency is negligible.

 \subsection{\texorpdfstring{Branching fraction of \( \psi(3686)\rightarrow{}hadrons\)}{Lg} }
 The uncertainty of the branching fraction of \(\psi(3686)\rightarrow{} \it hadrons\),   0.13\%  \citep{Luth:1975bh,BES:2002psd,ParticleDataGroup:2016lqr},
 is taken as a systematic uncertainty.

 \subsection{Total systematic uncertainty}
 The total systematic uncertainty for each $\psi(3686)$ data sample is obtained as the quadratic sum of all the systematic uncertainties.

 \section{Summary}
 By analyzing inclusive hadronic events, the number of $\psi(3686)$ events taken by the BESIII detector in 2021 is measured to be $(2259.3\pm 11.1)\times 10^6$, where the uncertainty is systematic and the statistical uncertainty is negligible. The numbers of $\psi(3686)$ events taken in 2009 and 2012 are also updated to be $(107.7\pm 0.6)\times 10^6$ and $(345.4\pm 2.6)\times 10^6$, respectively. Both are consistent with the previous measurements within one standard deviation, and a slight difference in the 2012 $\psi(3686)$ events relative to the previous one is caused by changing the off-resonance data from the previous $\tau$-scan data to the off-resonance data at $\sqrt{s}=$3.65 GeV in 2021.
 The total number of $\psi(3686)$ events for the three data samples is obtained to be $(2712.4\pm 14.3)\times 10^6$ by adding the above three yields linearly. This work provides an important parameter used in precision measurements of decays of the $\psi(3686)$ and its daughter charmonium particles.

 \acknowledgments{
 The BESIII collaboration thanks the staff of BEPCII and the IHEP computing center for their strong support.
 }


 \end{multicols}
 \vspace{-1mm}
 \centerline{\rule{80mm}{0.1pt}}
 \vspace{2mm}
 \begin{multicols}{2}

 \end{multicols}


\end{document}